\documentclass[%
 reprint,
 amsmath,amssymb,
 aps,
]{revtex4-1}

\usepackage{graphicx}
\usepackage{mathtools}
\usepackage{dcolumn}
\usepackage{bm}

\begin{document}


\title{\textit{``Should I break up with my girlfriend? Will I find another?"}
\\
Or: An Algorithm for the Forecasting of Romantic Options}


\author{Rashied B. Amini}
\altaffiliation[Also at ]{Caltech/Jet Propulsion Laboratory \\
Washington University in Saint Louis, Physics Department}
\email{rashied@nanaya.co}
\homepage[Visit: ]{www.nanaya.co}
\collaboration{\textit{on behalf of} Nanaya, LLC}
\affiliation{%
 Nanaya, LLC
}%

\date{January 7, 2015}

\begin{abstract}
The prospect of finding love may be scary but the prospect of committing to a relationship for the rest of your life is almost certainly scary. The secretary problem is a parallel to romantic decision making where an individual decides when to be satisfied with a selection choice in the face of uncertain future options. However, the secretary problem and its variations still do not provide a practical solution in a world where individual preference, goals, and societal context create a highly complex space of values that factor into decision making. In light of these complexities, we offer a general process that can determine the value of romantic options in a highly personal context. This algorithm is currently being developed into a service that will be available in 2015 for the general public.
\\
\\
\textit{Expressions and equations used and referenced in this release are for illustrative purposes and may not reflect the actual methodologies employed. Moreover, only some features and routines incorporated in the algorithm are described.}
\begin{description}

\item[Usage]

\end{description}
\end{abstract}

\pacs{Valid PACS appear here}
\maketitle


\section{\label{sec:level1}Introduction}

Online matchmaking services are extremely popular with about one third of marriages beginning with online dating \cite{pnas}.  Considerable effort has gone into in the development of personality tests that match people together based on results of psychometric assessment. While proprietary, there are many rudimentary methods available to perform matchmaking (\textit{e.g.} \cite{svd}). However, existing services for matchmaking often merely consider single users and identify potential matches between people who have not yet met.
\\
\\
While this serves as a viable means of introducing people who may be compatible, it neither offers the ability to determine long-term quality of relationship as a function of assessed personality and other system behavior nor the ability to weight a romantic option against other options. Inherent in the latter is the ability to  forecast the probability of meeting a person, whether online or offline, who may be considered a compatible romantic match. 
\\
\\
Work has been done to refine aspects of the secretary problem \cite{ferg} in a manner that provides additional clarity to its application in non-idealized contexts. For instance, this may include selecting subpopulations of a population \cite{vand}, human performance in solving the secretary problem \cite{lee}, discounting \cite{rasmussen}, weighting \cite{baba}, and many others.
\\
\\
In applying previous work, we must first consider new assumptions in our variation of the secretary problem. To do so, we define the following elements of the system: 

\begin{itemize}
  \item The \textit{user} is the subject, the individual making the romantic decision.
  \item	\textit{Groups} contain individuals that may or may not meet high-level demographic requirements set by the user. High-level demographic requirements include gender, age, \textit{etc}.
  \begin{itemize}
  \item \textit{Subgroups} exist and can be most simply understood as a subset of a whole group, or more precisely, the intersection of two or more groups (\textit{e.g.} gender, location, and ethnicity).
  \end{itemize}
  \item \textit{Traits} are quantitatively assessed values of an individual's personality.
  \item \textit{Suitors} are individuals within a group that meet the user's high-level demographic criteria. There are two categories of suitors:
  \begin{itemize}
  \item \textit{Failed matches} are suitors that the user will not entertain for a relationship due to mismatch of personality traits, lack of attraction, etc or suitors that will not accept the user for similar reasons. 
  \item	\textit{Matches} are suitors who fit the user's requirements and vice versa.
  \end{itemize}
  \item In cases where the user is in a relationship, a \textit{partner} is the individual with whom the user shares a relationship.
  \item \textit{Quality} is the user-based criterion by which we determine utility along any trait dimension. Though overly simplified, we can consider $ \mathbb{D}-$dimensional proximity of a suitor's traits from the user's ideal partner traits an example of quality. Though for psychological and relationship dynamics (not psychodynamic!) this example is undoubtedly an oversimplification.
\end{itemize}

\begin{figure*}[!hbt]
\centering
\includegraphics[width=\textwidth]{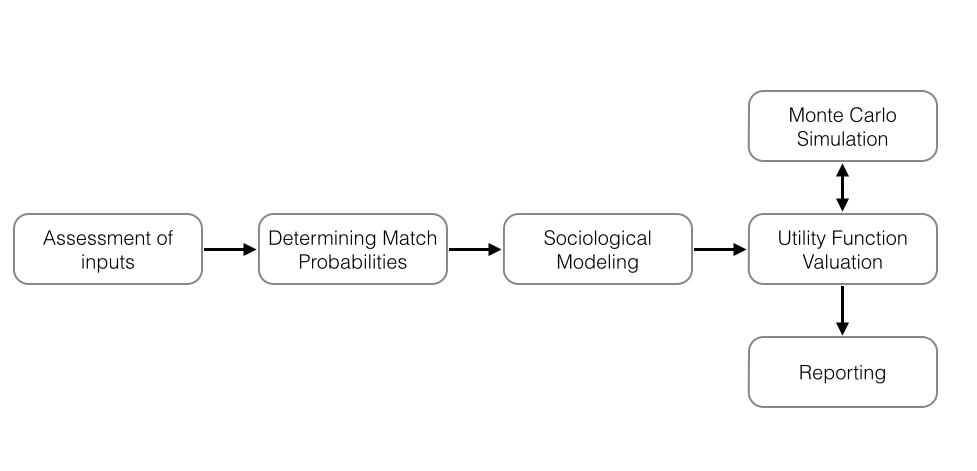}
\caption{The high-level process of the Nanaya algorithm.}
\label{figA}
\end{figure*}

With key system elements defined, we make the following assumptions that vary from existing solutions of the secretary problem:
\begin{enumerate}
\item A romantic partner may already exist which has its own unique value to the user.
\item User behavior, such as character and social behavior, is unknown \textit{a priori}.
\item There may exist more than one group that the user interacts with.
\begin{enumerate}
\item These groups are distinct and unique except in cases of an intersection of group that the user interacts with, creating a subgroup
\end{enumerate}
\item Information exists or can be meaningfully extrapolated from other information on the amount of suitors, failed matches, and matches within a (sub)group.
\item Information exists or can be meaningfully extrapolated from other information on the frequency of interaction between a user and groups with whom the user interacts.
\begin{enumerate}
\item Information exists or can be meaningfully extrapolated for bias of interaction of subgroups within known groups.
\end{enumerate}
\item Information regarding user preferences and romantic history is known from which reasonable extrapolations on romantic attraction and success can be made.
\item Values representing quality, user satisfaction, and/or stability of a relationship exist and
\begin{enumerate}
\item can be determined using utility functions which are themselves a function of constants, initial conditions, personality traits, \textit{etc}.
\end{enumerate}
\end{enumerate}
With these assumptions we see a contrast with previous variations of the Secretary problem. Notably, unknown user behavior and social context, incorporation of a value function for each suitor and match, dynamic system behavior, and the existence of significant quantified information of groups present a new variation of the Secretary problem.   Our internal studies have demonstrated that the available, closed-form solutions in literature are not relevant in all user cases. With little imagination, it is possible for the reader to reach a similar conclusion that closed-form solutions available are not always appropriate or descriptive.
\\
\\
In acknowledgement of the open-endedness of the problem as allowed by our assumptions, we have developed a numerical framework for simulating the availability and value of romantic options.  The Nanaya algorithm seeks to simplify a complex, highly dimensional problem in a manner that provides consistent results robust to varying user context. We further acknowledge that there is a limited ability to genuinely quantify initial conditions of social behavior, let alone personality, love, and emotions. The evolution of love and romantic options is truly chaotic - but Nanaya provides a sensible estimate.

\section{\label{sec:level1}Nanaya Algorithm Framework}
Notably, we suggest that what is presented in this paper is at the highest level a framework. We cannot offer explicit systems of equations nor acknowledge the level of detail actually incorporated due to sensitivity. We invite readers to use their imaginations as neccasry.
\\
\\
Our algorithm for forecasting long-term romantic futures must be interdisciplinary; \textit{i.e.} a framework that strictly utilizes mathematical and economic description without incorporating sociological and psychometric elements cannot provide any meaningful insights, let alone solutions. In our theoretical work, we determined many shared variables between psychometric results and factors used in sociological modeling and derived utility functions.
\\
\\
Figure \ref{figA} depicts the flow of the algorithm, not the user experience. We assume datasets or extrapolations exist or that allow for meaningful calculation of required secondary variables, \textit{e.g.} match probabilities.
\\
\\
\textit{Assessment of inputs} is the acquisition of user data such that all primary variables, e.g. psychometric test results, are obtained. In \textit{determining match probabilities} we utilize existing databases or extrapolations thereof to determine raw probabilities, $P_G(t)$, of finding matches of various quality across $G$ groups that the user interacts with. \textit{Sociological modeling} forecasts rates of interaction between a user and groups with which they interact based on which cumulative probabilities of finding a match can be ascertained. \textit{Utility function valuation} calculates the utility as a function of time for the user for any given relationship and the utility of strictly being alone. To simulate potential future matches we generate a pseudorandom set of suitors via Monte Carlo simulation. All results can be later reported. The following subsections further detail these processes.

\subsection{\label{sec:level2}Assessment of Inputs}
All required inputs are entered by the user. These inputs cover user psychometric assessment that assesses user personality and desired traits in a partner, populations to which the user belongs, their history in these groups, and emotions about the user's current relationship, if applicable. 

\subsection{\label{sec:level2}Determining Match Probabilities}
From user inputs, we can determine the subgroups that a user interacts with from which suitors exist. User desired partner traits are determined in terms of a \textit{window of compatibility} for each desired trait, \textit{i.e.} a range of trait values consistent with the user personality self-assessment. From subgroups and windows of compatibility we can derive the single encounter probability of finding a match. Single encounter probability is the chance that on any given encounter, the person in a (sub)group may be a romantic match for the user. It is found by searching within the database for all users within desired subgroups by the user across all trait dimensions. All matches falling within all windows of compatibility can also be further constrained based on proximity with the mean of the windows. This proximity can be considered a possible measure of match quality with more distant matches being of less quality.
\\
\\
In the event a subgroup is not statistically significant its definitions will be varied in a self-similar manner as a function of the most sensible local topographies in the dataset.
\\
\\
Probabilities are also determined in time based on direct user input and personality self-assessment to adjust for varying interest in partners in time. For instance, we consider the effect of increasing or decreasing sizes of windows of compatibility in time to account for romantic desperation or neuroticism in time. Forecasted demographic shifts in time at static locations or demographics patterns can also be incorporated. This will vary the single encounter probability of finding a match in time.
\\
\\
In contrast, Here we can consider a method employed in the literature as an example \cite{peter} for finding ``love." While we insist that ``love" is something wholly incalculable due to its nature, the probability of finding romantic matches is more appropriate. Moreover, we note significant gaps in \cite{peter} in the framework for determining single encounter probability how population bias is treated. Rather than relying on Drake's Equation, we take a first principle's approach to solving the problem of determining single encounter probabilities. This allows us to be as inclusive or exclusive as possible in defining populations while simultaneously noting the biases that exist in demographics. This was subjectively quantified in \cite{peter}. 

\begin{figure*}
\centering
\includegraphics[width=\textwidth]{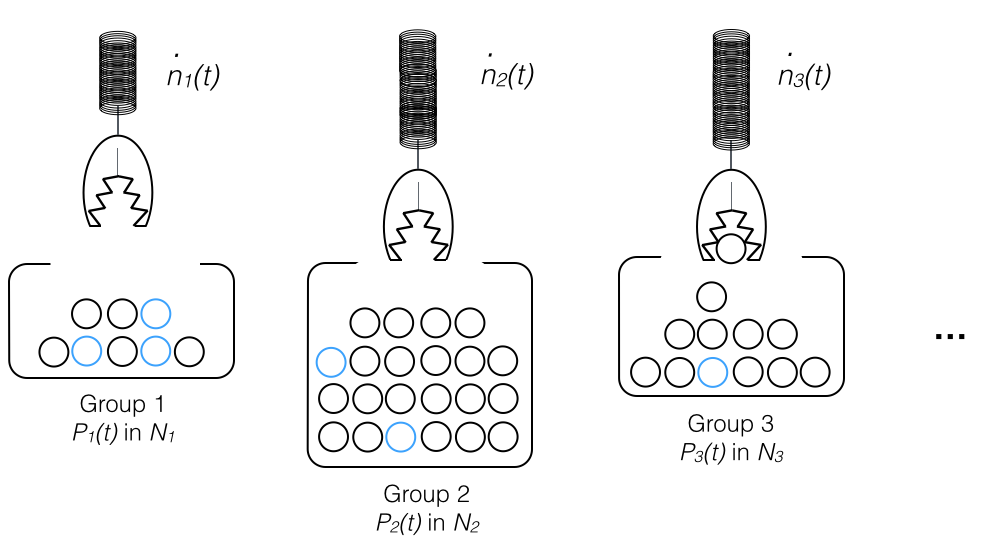}
\caption{A simple toy model for determining the cumulative probability of finding a match as an urn model. Each urn has a certain total number of balls, of which some fraction,$P$ are desired. A mechanical arm is lowered to draw a ball at a rate $\dot{n}$. 
\\
\\
In the real life case, each group of size $N_{pop}$ has its own single draw probability, $P_{pop}(t)$, as a function of time along with draw frequency, $\dot{n}_{pop}(t)$. Here we assume that replaceability is irrelevant as $N_{pop}>>1$. Moreover, the illustration does not incorporate mechanisms for demographic shift within (sub)groups; we acknowledge this exists.}
\label{figB}
\end{figure*}

\subsection{\label{sec:level2}Sociological Modeling}
\begin{figure}
\centering
\includegraphics[width=0.48\textwidth]{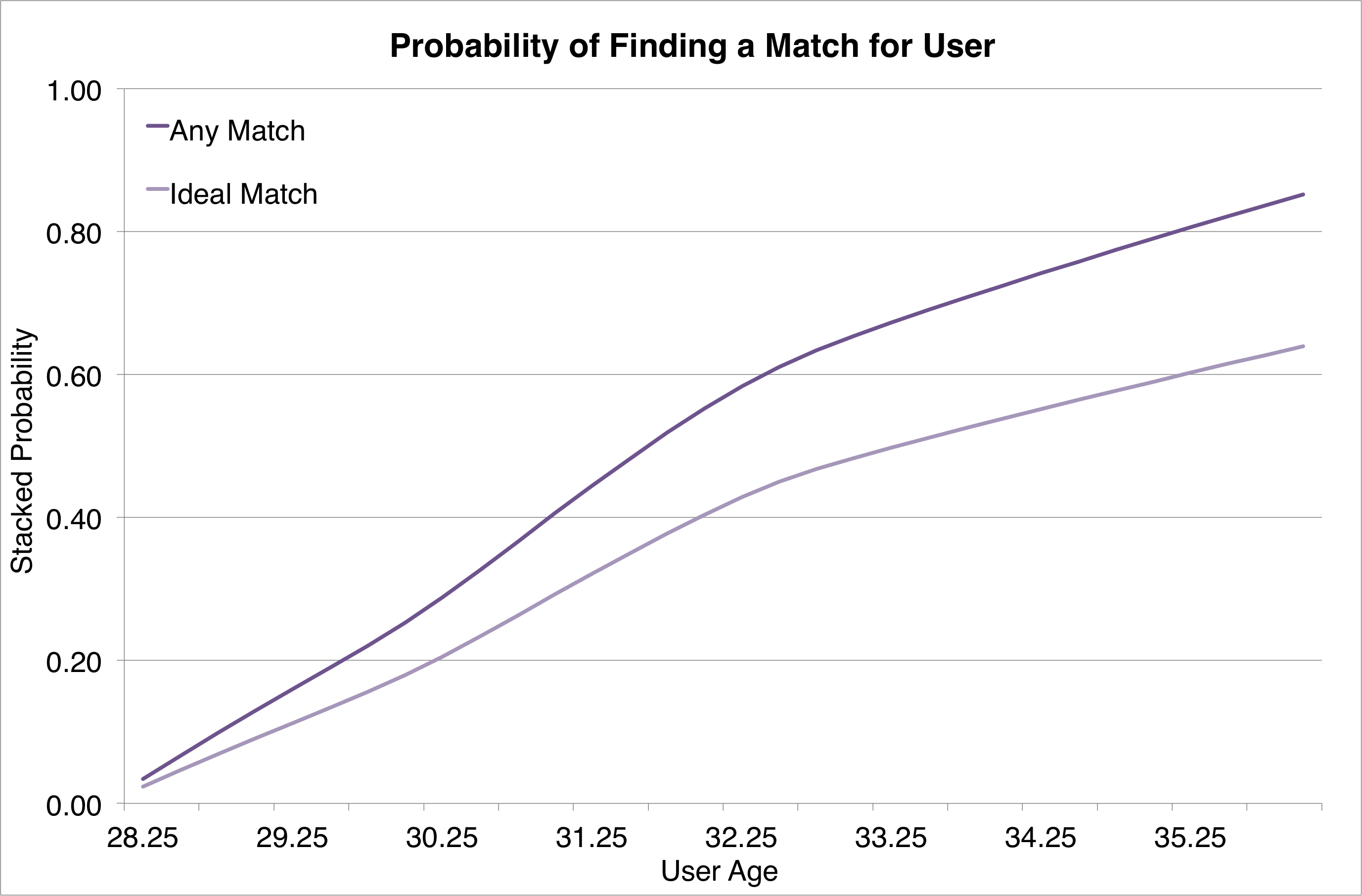}
\caption{The probability in time of finding a match for a 28 male year old user as broken down in terms of quality. We define \textit{ideal} to be of some arbitrary criterion, a function of quality, which is exclusive of \textit{all} matches. For example, we note that the interactions of the user in this case study make it likely that most matches will be of ideal quality.}
\label{chartA}
\end{figure}

\begin{figure}
\centering
\includegraphics[width=0.48\textwidth]{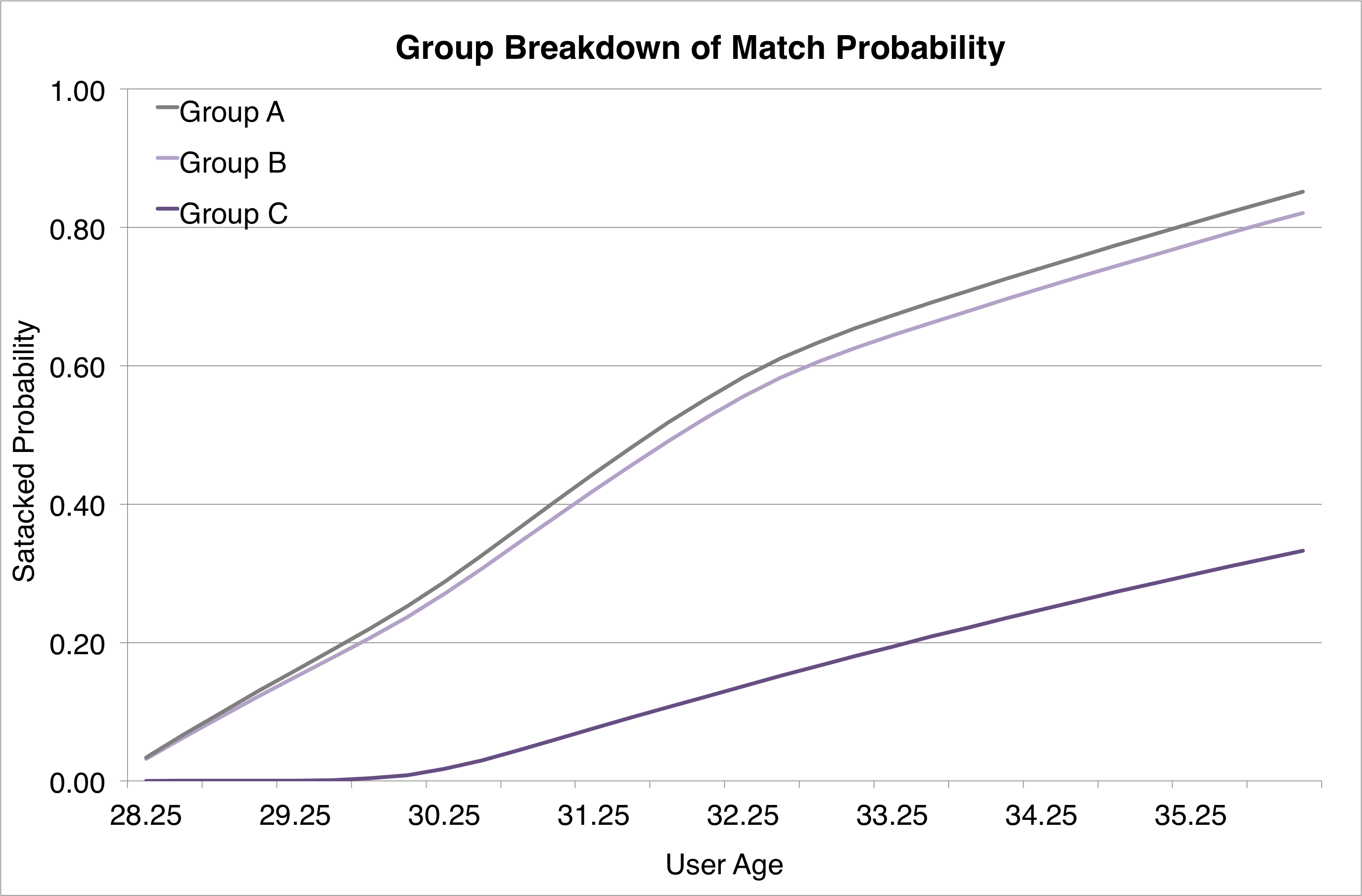}
\caption{The probability in time of finding a match for a 28 male year old user as broken down in terms of (sub)group of interaction.}
\label{chartB}
\end{figure}

Without knowing the number of people a user interacts with, the single encounter probability of finding a match is useless. Therefore it is vital to estimate the number of people a user interacts with within each (sub)group to determine the probability of finding a partner. 
\\
\\
We maintain such an estimate of frequency of interaction is calculable and, minimally, a function of user personality, population characteristics such as demographics and size, and user social history within the group. Such estimates are approximate and approximately verifiable through empirical study. Several functional forms of the model were studied for accuracy and exist for different personality types and social circumstances.
\\
\\
Each (sub)group the user interacts has its own time varying single encounter probability and frequencies of interaction. Using these, we determine a cumulative probability in time of finding a match. This problem is a modified urn problem Figure \ref{figB} illustrates the problem with a toy model. Notably, results from Nanaya may lead to investigating a variation of the Urn problem wherein ``mutation" occurs after a selection event. In our case, that would be personality shift as a result of a relationship prior to returning to the original population. 
\\
\\
For simplicity, in this paper we assume very large populations. At this limit we consider a binomial distribution for determining cumulative probabilities of finding matches. With such an approach we can determine weighted average single encounter probability across populations as a function of total encounters per group. Here, their time average single encounter probabilities are appropriate. We can then determine a cumulative probability as a function of time per group using the fractional probability over the normalized sum.
\\
\\
The complexity of the analytical solution increases in cases where groups depart from the large $N$ limit and in incorporation of user desire to revisit previously rejected and/or failed matches. 
\\
\\
The prototype for the Nanaya algorithm was designed for the large $N$ limit and static single encounter probabilities. Thus, we can consider the example of a user and their chances of finding a match in time in this simple case. The prototype integrates results of single draw probability determination for each (sub)group and uses the sociological modeling to determine probabilities of encountering a match. Figures \ref{chartA} and \ref{chartB} both show the same test case as a \textit{stacked} probability plot. In Figure \ref{chartA} we consider the probability of finding a partner in time as distinguished by quality while in Figure \ref{chartB} we distinguish by the group from which a match is likely found.

\subsubsection{\label{sec:level2}Utility Function Valuations}
\begin{figure}
\centering
\includegraphics[width=0.48\textwidth]{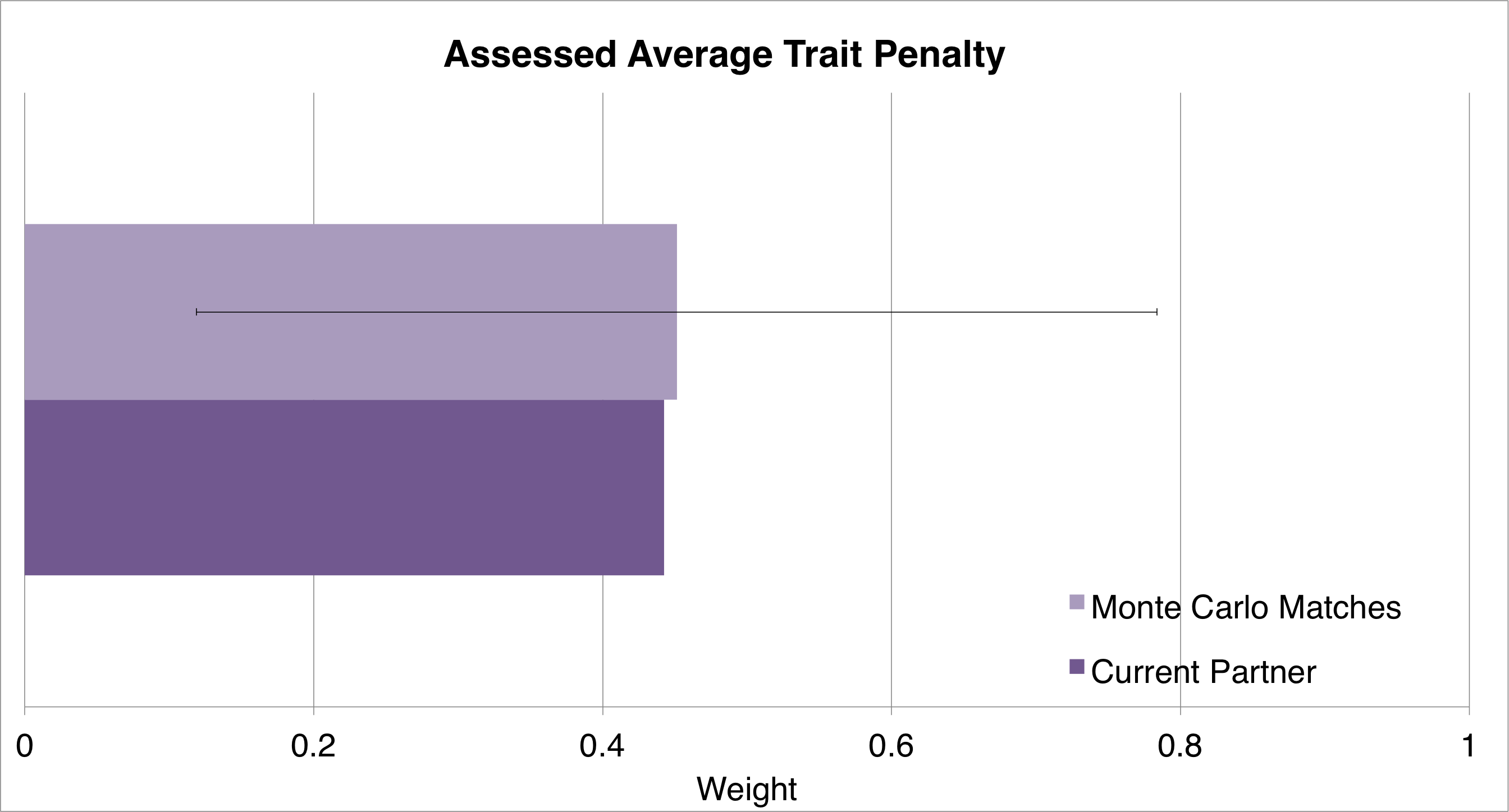}
\caption{The averaged penalty (\textit{e.g.} $\bar{w}_{g,i}$ across all traits and all Monte Carlo simulation generator suitors as compared to the average penalty of a notional girlfriend.}
\label{chartC}
\end{figure}

The ultimate purpose of the Nanaya algorithm is to determine the value of romantic options available. We consider there to be three different options: to remain in an existing relationship, to be single with no intention of being in a relationship, and to be single and interested in a relationship of opportunity. To calculate the utility of these options, we consider two functionals. We define functionals to calculate a value of relationships between two individuals and a functional for the value of being single. The units of utility are arbitrary; it is in comparison of values between forecasts that we can determine optimal outcomes.
\\
\begin{equation}
\prod_g^2{\sum_i^\mathbb{D}{a_{g,i} \mathrm{e}^{-w_{g,i} \mid trait_{2,i}-trait_{1,i} \mid t}}}
\label{eq1}
\end{equation}
\\
The value of a relationship between two individuals is derived from a functional that operates in a highly dimensional personality space that is a function of the traits measured in the psychometric assessment and other values input by the user. Exactly, we assert $\mathbb{D}>3$ is fundamentally required due to psychological aspects of the problem. Weighted proximity between median user trait desires and partner trait values for both partners for a given trait indicates value of the relationship for that given trait. Equation \ref{eq1} is an example of such a function. Naturally, this value is a function of time wherein a relationship depreciates overtime. In Equation \ref{eq1} we can tell that relationships depreciate in a compound manner over time.
\\
\\
A system of differential equations over personality space that converges on an equation like Equation \ref{eq1} may also be constructed which makes it possible to evaluate system stability in time. Naturally, in a subjectively defined, highly dimensional space it may be all but impossible to draw fundamental insights or conclusions - but it would be really cool. \cite{myBlog}
\\
\\
Values for specific individuals can be input for such a utility function. In evaluating the utility of \textit{possible} matches we can generate a pseudorandom set of suitors through a Monte Carlo simulation. The personalities of the pseudorandom suitor set are determined by principle component analysis from the occupied personality space of the subgroups accessible to the user. Therefore we evaluate relationships that are most probable, not most ideal. This provides  realistic forecast as opposed to a wholly randomized one or one that is overly ideal. We can consider the average and standard deviation of the penalty, $w_{g,1}$, assessed to pseudorandom suitors as compared to a specific individual. In Figure \ref{chartC} we compare to a notional girlfriend. 
\\
\\

\begin{figure}
\centering
\includegraphics[width=0.48\textwidth]{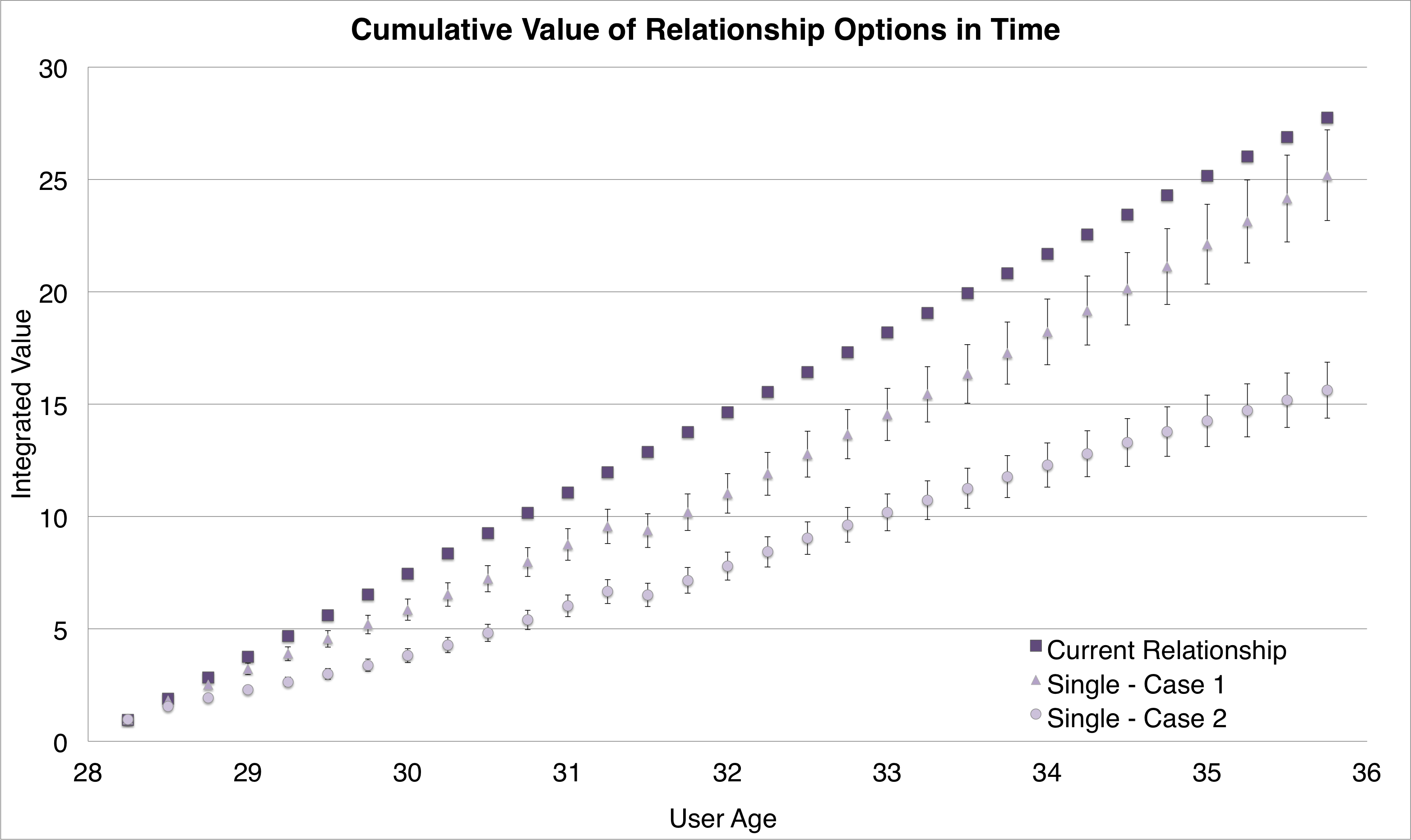}
\caption{The romantic options available, and their value in time, for the on-going example of a 28 year old man. The forecasted value of \textit{Single - Case 1} and \textit{Single - Case 2} are based on two different locations with nearly identical social interaction.}
\label{chartD}
\end{figure}

The form of the functional of being single is based on user personality and direct user response to select questions that indicate user life goals. This indicates what ``single space" looks like.
\\
\\
With these two building blocks and through incorporation of cumulative probabilities for each subgroup as determined in the prior step, we can determine the optimal decision for the user given highly personal user context. In all forecasting, errors and other sources of uncertainty are propagated. In Figure \ref{chartD} we forecast the value in time of several relationship options; one is the existing relationship while the other two are different cases of being single based on location. In this example, it is clear that the user should remain with his girlfriend despite good odds in Figures \ref{chartA} and \ref{chartB} of finding another.

\begin{figure}
\centering
\includegraphics[width=0.48\textwidth]{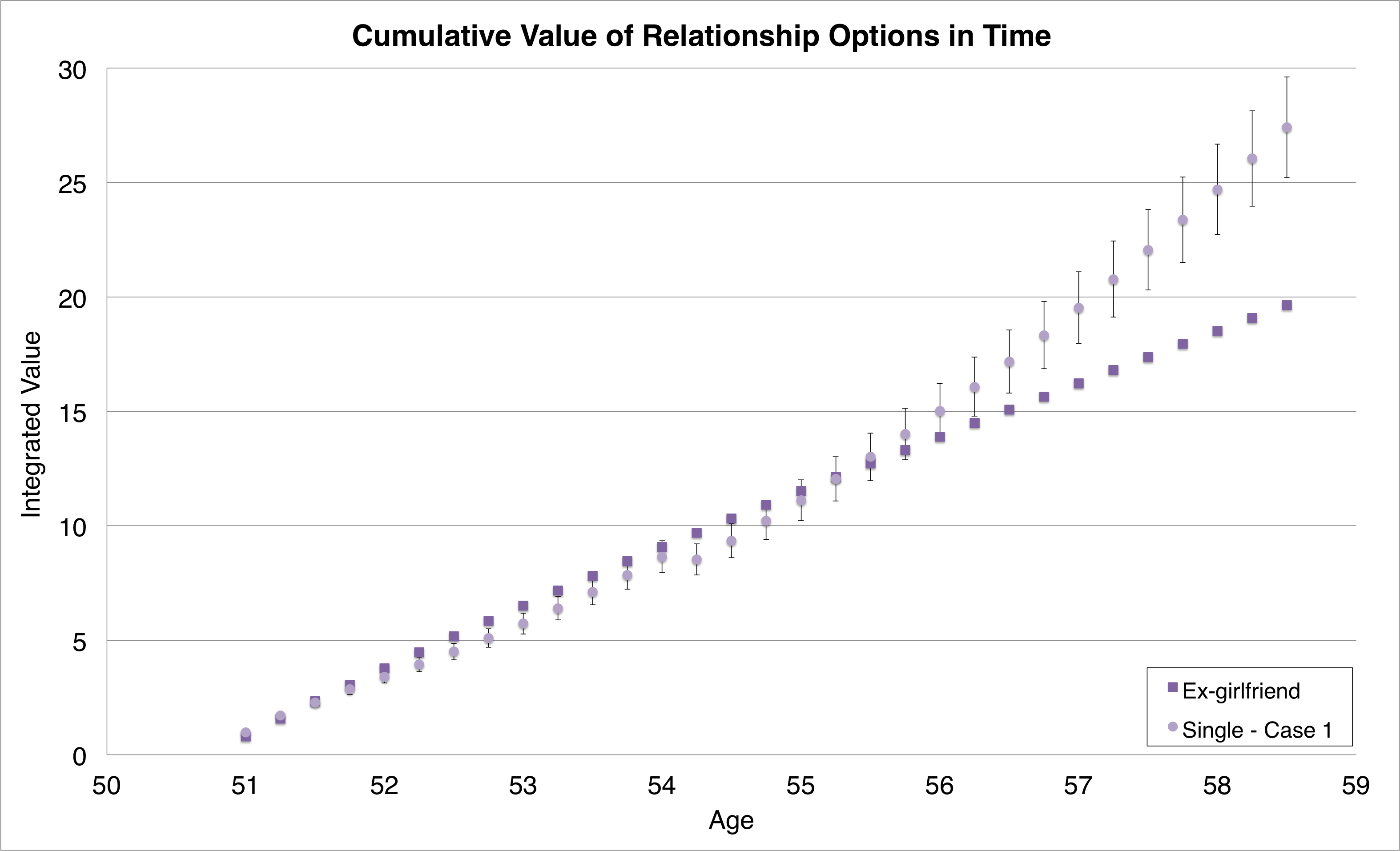}
\caption{A new case of a 51 year old single man wherein a previous romantic decision was reaffirmed by algorithm output.}
\label{chartE}
\end{figure}

In Figure \ref{chartE} we consider the new case of a 51 year old man who is currently single having left a relationship in the past. If he were to repeat the relationship, the forecast predicts greater utility in remaining single per his social circumstances. 

\subsection{\label{sec:citeref}Reporting}
The values determined with the utility functions and cumulative probabilities can be reported to the user for decision making. Additional synthetic metrics or comparison to the existing dataset of users (this need not be personal) can indicate scores of the following to the user
\begin{itemize}
\item Their romantic selectivity 
\item Their opportunity for non-romantic social growth
\item Their opportunity to find a match or love, however that may correlate
\item If they have a partner:
\begin{itemize}
\item The quality of their partner as compared to non-failed suitors
\item Whether remaining in a relationship or returning to being single will probably provide maximum utility
\end{itemize}
\end{itemize}

Moreover, the data sets that are used to derive single encounter probabilities are also useful in determining ideal subgroups for the user. Results from database queries can be used to determine ideal and worst locations, jobs, demographics, and other subpopulations to best find friends or lovers.

\section{\label{sec:level1}Future Work}
Validation is one of the trickiest parts of long-run forecast models. Thus far, validation efforts have focused on those who have been felt firm in their previous romantic decisions, ideally for several years. This poses issues in the evolution of personalities through relationships as well as remembering previous social circumstances. We look forward to on-going validation efforts and methodologies that allow us to reduce the amount of input necessary to the model. 
\\
\\
A remaining issue is incorporation of bisexuality into the algorithm. Non-binary orientation creates considerable complexity as bisexual individuals may not seek identical traits in partners of both genders or value both genders equally. This impacts windows of compatibility, thus single encounter probabilities, and also empirically descriptive utility functions. Therefore theoretical work and continued validation from satisfied bisexuals remains necessary.
\\
\\
A extension for polyamorous relationships is also under development. Needless to say, moving from a $2-$ to a $3-$ or $N-$body problem is quite difficult informationally and computationally. Naively, from a non-linear systems perspective we suspect long-term polyamorous outcomes will be chaotic and generally unstable.
\\
\\
Another feature that was intentionally overlooked is the incorporation of intertemporal choice. While a realistic description would incorporate this phenomena, and indeed factors of diminishing utility can be readily incorporated, we believe it to be more philosophically sound to give \textit{prescriptive} as opposed to \textit{descriptive} assessment of utility as to assure the user that patience may bring better outcomes rather than hasty decision making. Nonetheless, validation can only be done through \textit{description} of past events and incorporation of diminishing utility. Herein, previous romantic experiences need to be assessed along side psychometric evaluation \cite{intertemp98}. Therefore, we look forward to incorporating this diminishing utility in future model validation.

\section{\label{sec:level1}Summary}
We can summarize the process with the following steps:
\begin{enumerate}
\item User romantic and life desires, social behavior, and feelings toward an existing partner are entered into the algorithm.
\item The probabilities of finding a match across all parts and times of the users life are calculated.
\item Social behavior across all groups a user interacts or will interact with are modeled.
\item The above data are entered into a utility function equation which is scaled to account for time spent while single.
\item Results are reported to the user in an easy to understand format.
\end{enumerate}

\section{\label{sec:level1}Conclusion}
Thus, with such an algorithm it is possible to numerically solve one of mankind's most timeless problems in a consistent and objective manner. \textbf{Colloquially, we can determine the chances of finding love and where and when it will be found. We can also determine if any given relationship is worth the time and potential emotional investment.} 
\\
\\
For individuals who are not self-certain and lack the access to impartial help, Nanaya provides clarity to romantic and social decisions in an unbiased manner. For those who are in the process of making difficult decisions, Nanaya brings the sort of objective affirmation that is otherwise impossible. For others, including the author, Nanaya is a unique application of systems analysis and mathematical methods that provides remarkably self-consistent and personal results. Beyond the algorithm significant data is obtained and is then accessible which can indicate ideal cities, careers, ethnicities for finding friendship or romance through database queries. 
\\
\\
We must emphasize the use of ``colloquially." Love is more than just a number. Despite consistency and objectivity, we believe there is far more to human relationships than can be put to numbers. In highly dimensional problems with unverifiable initial conditions, it is all but impossible to predict an exact outcome. The Nanaya algorithm reduces a \textit{difficult mathematical problem} into straight-forward algorithm with practical results. The point of Nanaya is to bring certainty and affirmation to decision making - not to make a decision for the user. Moreover, we believe that the input process itself may be just as valuable for the user than the model results. For many individuals on the cusp of romantic decisions, many questions about life goals and romantic desires have never been formally or externally broached. Nanaya provides a neutral and objective medium to ask these questions with minimal priming. 
\\
\\
For more information and use, please see \url{www.nanaya.co}

\section{\label{sec:level1}Acknowledgements}
I'd like to thank the Nanaya team, especially C\'esar Andr\'es Liz\`arraga. Additionally, I'd like to thank Saeid Amini (GWU), Hosam Mahmoud (GWU), and Robert Shishko (Caltech/JPL).

\end{document}